\documentclass[lettersize,journal]{IEEEtran}
\usepackage{amsmath,amsfonts}
\usepackage{algorithmic}
\usepackage{algorithm}
\usepackage{array}
\usepackage{textcomp}
\usepackage{stfloats}
\usepackage{url}
\usepackage{verbatim}
\usepackage{graphicx}
\usepackage{cite}
\usepackage[font = small, labelsep = period]{caption}

\usepackage{xcolor}
\usepackage{authblk}

\begin{document}

\title{Characterization of Asymmetric Gap-Engineered Josephson Junctions and 3D Transmon Qubits}

\author{Zach Steffen, S. K. Dutta, Haozhi Wang, Kungang Li, Yizhou Huang, Yi-Hsiang Huang, Advait Mathur, \mbox{F. C. Wellstood}, B. S. Palmer %

\thanks{ }
\thanks{Zach Steffen, Haozhi Wang, Yizhou Huang, Yi-Hsiang Huang, Advait Mathur, and B. S. Palmer are with the Quantum Materials Center at University of Maryland, College Park, MD 20742, USA and also with Laboratory for Physical Sciences, College Park, MD 20740, USA (e-mail: \mbox{zsteffen@lps.umd.edu}; hzwang01@lps.umd.edu; yhuang@lps.umd.edu; yhhuang@lps.umd.edu;  bpalmer@umd.edu).}%
\thanks{Kungang Li, S. K. Dutta, and F. C. Wellstood are with the Quantum Materials Center at University of Maryland, College Park, MD 20742, USA (e-mail: kungang@umd.edu; sudeep@umd.edu; well@umd.edu).}%
\thanks{ }}%
%
\markboth{ }%
{Shell \MakeLowercase{\textit{et al.}}: A Sample Article Using IEEEtran.cls for IEEE Journals}

\IEEEpubid{ }

\maketitle

\begin{abstract}
We have fabricated and characterized asymmetric gap-engineered junctions and transmon devices. To create Josephson junctions with asymmetric gaps, Ti was used to proximitize and lower the superconducting gap of the Al counter-electrode. DC IV measurements of these small, proximitized Josephson junctions show a reduced gap and larger excess current for voltage biases below the superconducting gap when compared to standard Al/AlO$\mathrm{_x}$/Al junctions. The energy relaxation time constant for an Al/AlO$\mathrm{_x}$/Al/Ti 3D transmon was $T_{1}= 1\ \mu s$,  over two orders of magnitude shorter than the measured $T_1 = 134\ \mu s$ of a standard Al/AlO$\mathrm{_x}$/Al 3D transmon. Intentionally adding disorder between the Al and Ti layers reduces the proximity effect and subgap current while increasing the relaxation time to $T_1 = 32\ \mu s$. 
\end{abstract}


\section{Introduction}
\IEEEPARstart{Q}{ubits} based on superconducting circuits are one of the leading platforms in implementing a quantum computer \cite{sc_circuits_and_qi,Kjaergaard_review, quantum_engineers, google_roadmap}. 
One limitation of superconducting qubits is their coherence times which places a hard limit on  fidelity of operations and time of computation in the absence of error correction. While state-of-the-art transmon qubits have energy relaxation times reaching up to $T_1 = 500$ microseconds \cite{Place2021, Wang2022}, significant further improvement is required for most quantum computation applications. 

In a transmon qubit, intrinsic losses come predominantly  from the coupling of the qubit to two-level systems  in lossy dielectrics \cite{TLS_1, TLS_2, interface_loss} and non-equilibrium quasiparticle loss channels \cite{ne_qp1,   fluxonium_neqp, IBM_neqp, UW_neqp}. 
Quasiparticles in superconducting qubits can tunnel across the Josephson junction, absorbing a photon, and relaxing the qubit. The decay rate from the qubit's excited state $|e\rangle$ to the ground state $|g\rangle$ as a result of quasiparticles tunneling across the qubit junction can be written as 
\begin{align}
    \Gamma_{e\rightarrow g} 
    &=\frac{E_C}{hf_{ge}e} \Big(I_{ \rightarrow }(f_{ge}) + I_{ \leftarrow }(f_{ge})\Big).
\end{align}
%
Here  $E_C$ is the charging energy, $h$ is Planck's constant, $e$ is the electron charge, and $(I_{ \rightarrow }(f_{ge})+I_{ \leftarrow }(f_{ge}))$ is the sum of the quasiparticle current in both directions across the junction at a bias equivalent to the qubit frequency $f_{ge}$ \cite{ROGOVIN1974}. $I_{\rightarrow}/I_{\leftarrow}$ is determined by the specifics of the Josephson tunnel barrier and the density of quasiparticles  $n_{qp}$ at the junction. 
For a BCS superconductor in thermal equilibrium, $n_{qp}$   decreases exponentially with decreasing temperature ($T$) \cite{tinkham} resulting in a reduced quasiparticle current and therefore reduced $\Gamma_{e\rightarrow g}$. While theoretically, Al below  $T < 50$ mK should have $n_{qp} < 10^{-12} \mathrm{\mu m}^{-3}$; experiments have reported nonequilibrium $n_{qp} \sim 0.1 - 1\ \mathrm{\mu m}^{-3}$ \cite{deVisser_qp_dist, aumentado_ge, Sun_qp_dist, Riste2013, kyle_neqp, Bal_qp_dist}.

Recently, $T_{1}$ measurements of 3D transmons have shown an increase in $T_{1}$ as $T$ was increased from 20~mK to 150~mK. To explain this anomalous temperature dependence, the superconducting gaps $\Delta_{1,2}$ for the two Al layers forming the Al/AlOx/Al junction were modeled to be slightly different, such that for low $T$, the density $n_{qp}$ at the junction increases in one of the electrode layers, thus resulting in a decrease in $T_1$\cite{zhang_2020}. This model, which is distinct from quasiparticle trapping models,  suggests that increasing the  $|\Delta_1 - \Delta_2| > hf_{ge}$ should increase $T_{1}$, even in the presence of nonequilibrium quasiparticles. Recent results by oxygen-doping Al have shown long lived transmons informed by these predictions \cite{Li2022}. Other work has also discussed the improvement in the performance of similar devices by gap-engineering the junction \cite{aumentado_ge,yamamoto_2006,giampiero}.
\IEEEpubidadjcol

With this motivation, we have fabricated and characterized  junctions and transmon qubits with asymmetric superconducting energy gaps across the junction. In particular, the superconducting gap of the junction's larger-volume top electrode was lowered via the proximity effect by evaporating Ti on top of Al. Because the superconducting energy gap of Ti is lower than that of Al, the effective superconducting order parameter of an Al/Ti bilayer will have a value less than that of an Al film. 

In Section \ref{setup}, we describe the fabrication of the Al/Ti bilayer with and without an additional thin AlO$\mathrm{_y}$ disorder added in between these layers.
The theoretical critical temperature depends not only on the order parameters and coherence lengths of the two films but also on the AlO$_y$ interface barrier resistance \cite{Nb_proximity, Fominov_2001, Zhao_2018, kyle_2019}. 
DC characterization of these junctions is discussed in Section \ref{sec:dctunn}. Junctions with the Ti capping layer directly on the Al display a reduced energy gap but also a significant amount of tunneling current in the subgap region. We model  the IV's, including some of subgap features, using a multiple-Andreev reflection (MAR) theory and extract junction parameters. Lifetime measurements of 3D transmons with these electrode structures are discussed in Section \ref{sec:lifetime}. In particular, the Ti capping layer displays a substantial shorter lifetime of $T_1 = 1 \ \mu \mathrm{s}$ when compared to  standard Al/AlO$_{x}$/Al junctions while introduction of AlO$\mathrm{_y}$ disorder between the Al and Ti layers can partially restore the qubit $T_{1}$.

\section{Experimental Setup} \label{setup}

\begin{figure}
    \centering
    \includegraphics[width = 3.4in]{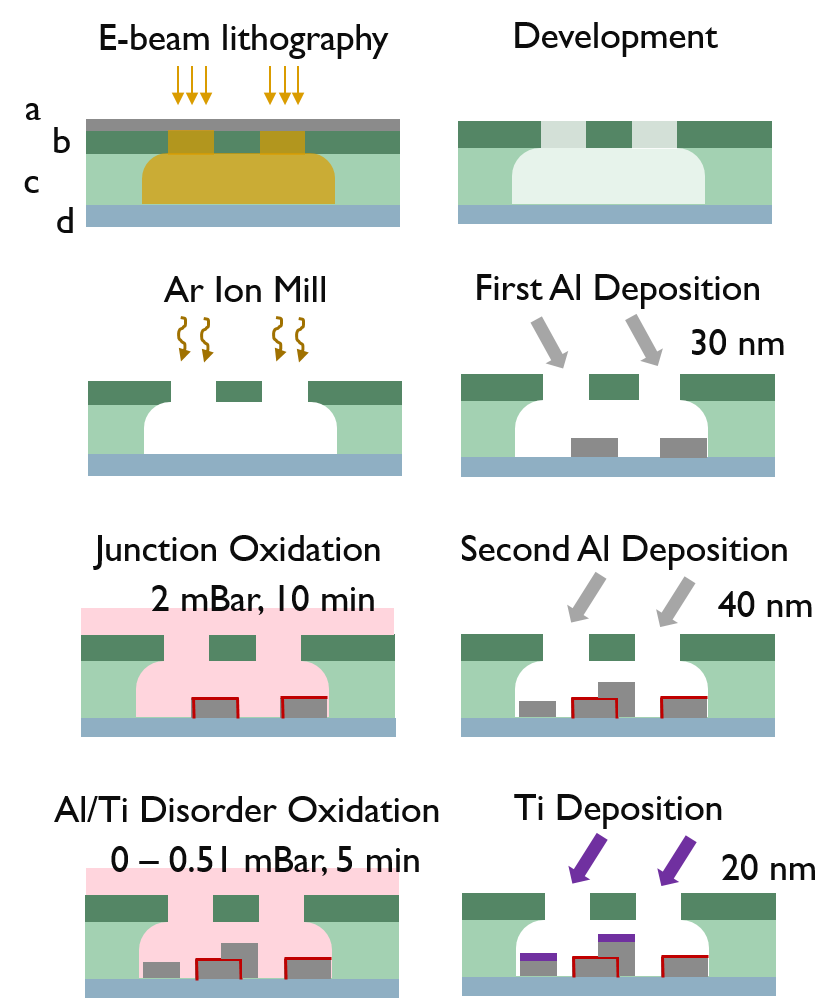}
    \caption{Schematic of the Dolan bridge fabrication process for asymmetric junctions.  We begin the process by spinning and baking   MMA (c), and ZEP (or PMMA for the Manhattan junctions) (b), followed by depositing a 10 nm  anti-charging Al layer (a) on  a silicon or sapphire substrate (d). The desired  pattern is defined via electron-beam lithography then developed. Within the evaporator, an Ar ion mill cleaning removes residual resist. Next, the junction is formed in standard double-angle evaporation where Al is deposited at the first angle; oxygen is let into the chamber to form the tunnel barrier; then Al is deposited at a second angle. To form the superconducting bilayer, we optionally perform another oxidation at a lower dose to create a layer of disorder. Finally, Ti is deposited at the same second angle, forming the bilayer directly on top of either the second Al or AlO$_{\mathrm{y}}$ disordered layer}
    \label{fig:fab}
\end{figure}

For this work, a variety of small area ($\sim$ 150~nm x 150~nm) low critical-current junctions were fabricated either by the Dolan bridge \cite{Dolan1977} or Manhattan technique \cite{manhattans}. To fabricate the junctions (see Fig.~\ref{fig:fab}), a lift-off stencil  was created using an electron-beam lithography system in a bilayer electron-beam resist with an undercut allowing metal to be deposited at an angle on the substrate.

After developing the resist, multiple metal depositions and oxidations were performed  either in a Plassys MED 550 S electron-beam evaporator or using a tungsten wound basket in a turbo-pumped thermal evaporator. For each junction, 30~nm of Al was first deposited, followed by a static oxidation with a fixed dose of 0.8 - 3 mBar  O$_{2}$ for 10 minutes to form the insulating tunnel barrier. Following the first oxidation, the chamber was evacuated and a secondary metal counter-electrode was deposited at a different angle. The composition of this top electrode was 50~nm Al for the standard Al/AlO$_{\mathrm{x}}$/Al  junctions, 45~nm Al followed by 5~nm of Ti for the Al/Ti bilayer, or 40~nm Al followed by an additional oxidation and a deposition of 20~nm Ti for the disordered Al/AlO$_\mathrm{y}$/Ti bilayer. The Al films were deposited at a rate of 0.5~nm/s; the Ti films were deposited at 0.2~nm/s. 

Next, the extraneous metal on the top of the resist was lifted off in a hot bath of NMP. Table \ref{tab:table1} has a summary of the devices that were fabricated in this study. Because the contact pad areas were approximately $10^7$ times larger than the junction areas, we argue each of these devices is effectively a single Josephson junction with asymmetric electrodes.

\begin{table}
\begin{center}
\begin{tabular}{| c  c  c  c |}
\hline
Junction & $O_2$ disorder & $\Delta_2$ & $R_n$\\
Composition & (mbar$\cdot$min) &  $(\mu eV)$ & $(k\Omega)$\\
\hline
Al/AlO$\mathrm{_x}$/Al& - & 200 & 18.6 \\
\hline
Al/AlO$\mathrm{_x}$/Al& - & 200 & 8.3 \\
\hline
Al/AlO$\mathrm{_x}$/Al/Ti& - & 100 & 6.3 \\
\hline
Al/AlO$\mathrm{_x}$/Al/Ti& - & 140 & 7.0 \\
\hline
Al/AlO$\mathrm{_x}$/Al/Ti& - & 110 & 8.5 \\
\hline
Al/AlO$\mathrm{_x}$/Al/AlO$\mathrm{_y}$/Ti& 0.3 & 120 & 2.6 \\
\hline
Al/AlO$\mathrm{_x}$/Al/AlO$\mathrm{_y}$/Ti& 0.6 & 230 & 8.8 \\
\hline
Al/AlO$\mathrm{_x}$/Al/AlO$\mathrm{_y}$/Ti& 0.6 & 220 & 8.5 \\
\hline
Al/AlO$\mathrm{_x}$/Al/AlO$\mathrm{_y}$/Ti& 1.8 & 220 & 9.0 \\
\hline
Al/AlO$\mathrm{_x}$/Al/AlO$\mathrm{_y}$/Ti& 2.8 & 220 & 8.5 \\
\hline

\end{tabular}
\caption{Comparison of measured superconducting gap for the top counter-electrode $\Delta_{2}$  and normal state resistances for different junction compositions. The counter-electrode gap $\Delta_2$ is extracted from the quasiparticle-quasiparticle tunneling peak assuming the bottom electrode $\Delta_1~=~190~\mathrm{\mu eV}$. Adding Ti to the counter-electrode reduces $\Delta_2$ while adding an AlO$_y$ barrier greater than 0.3 mbar$\cdot$min between the Al and Ti layers restores $\Delta_2$.}
\label{tab:table1}
\end{center}
\end{table}

To perform low-temperature measurements, the devices were mounted and cooled on a cryogen-free Leiden  dilution refrigerator with a base temperature below 20 mK. To perform the junction DC measurements, the device was glued using GE varnish to a copper package and then wire-bonded using Al wire to four separate leads on a PCB with either a RC low-pass, LC low-pass filter, or no filters. Each  measurement line then went through a combined  pi-powder low-pass filter on the mixing chamber plate before going to room temperature using thermal coax cables. Four-wire differential conductance measurements were performed by applying a combined  DC and low frequency 13~Hz AC voltage excitation and measuring the AC current with a lock-in amplifier. Using the additional two wires across the junction, the voltage was amplified with a PARC pre-amplifier before being digitized or measured with an additional lock-in amplifier. After the data were acquired, the dI/dV measurements were  numerically integrated resulting in the IV curves presented in the next section.

To perform qubit energy relaxation measurements, a transmon qubit was fabricated on a sapphire substrate and mounted in a 3D rectangular cavity machined out of an aluminum-6063 alloy. The  bare resonant frequency of the cavity was approximately at 7.8~GHz and a loaded quality factor, limited by the output coupler, of $Q_{L} \simeq 12,000$. The cavity was mounted in two Amumetal 4K magnetic shields on the mixing chamber plate of the same Leiden dilution refrigerator. To measure $T_{1}$, a $\pi$-pulse was used to initially excite the qubit from the ground state, followed by a variable delay before measuring the qubit state through dispersive cavity readout \cite{Wallraff2004}. Measurements were averaged 500-5000 times. The total acquisition time of a single decay curve varied between one and ten minutes. $\overline{T_1}$ values are averaged from 10's of individual curves at base temperature.

\section{DC Tunneling Measurements} \label{sec:dctunn}
\begin{figure}
    \centering
    \includegraphics[width= 3.4in]{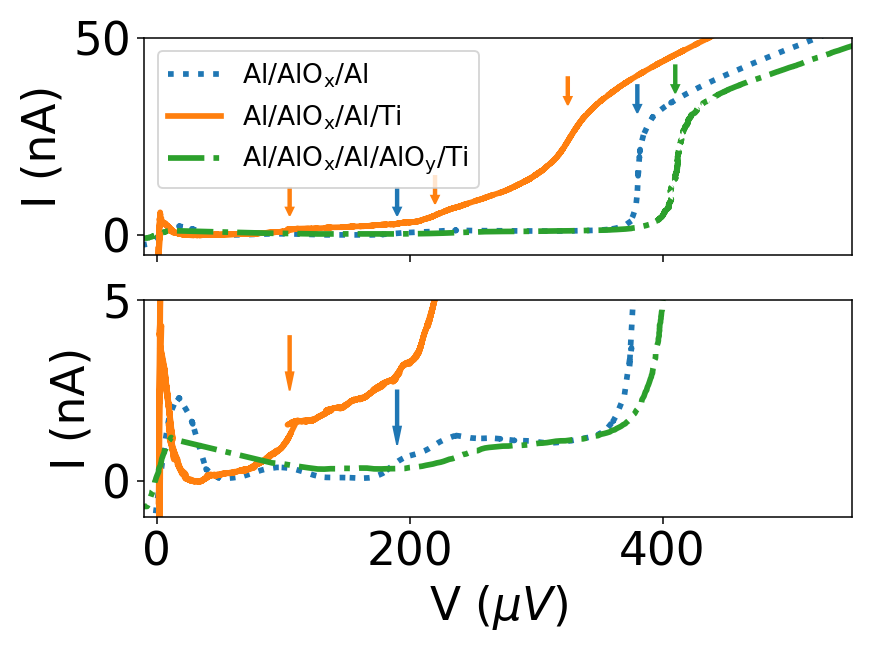}
    \caption{IV traces of three Josephson junctions at 20 mK with different counter-electrode structures (bottom plot is the same data as the top plot but with reduced current range). The  dotted trace is a standard Al/AlO$\mathrm{_x}$/Al transmon junction. 
    The solid trace is data for an Al/AlO$\mathrm{_x}$/Al/Ti junction (note reduction in $V$ to observe quasiparticle-quasiparticle tunneling due to the Al/Ti bilayer counter-electrode). 
    The dot-dashed curve is a measurement of a Al/AlO$\mathrm{_x}$/Al/AlO$\mathrm{_y}$/Ti junction with an additional AlO$_y$ disorder layer between the Al and Ti bilayer. Arrows mark visible steps at expected  $\Delta_1$, $\Delta_2$, or $\Delta_1+\Delta_2$ values.
    }
    \label{fig:IV_comparison}
\end{figure}

To quantify the superconducting gap electrodes and characterize the quality of the junction, DC measurements were performed. Fig. \ref{fig:IV_comparison} shows integrated IV curves for three different junctions: with no Ti (dotted), Al/Ti bilayer (solid), and with added AlO$_\mathrm{y}$ disorder between the Al and Ti layers (dot-dashed). These junctions were nominally identical to the measured transmon junctions. 

For the Al/AlO$_\mathrm{x}$/Al junctions, the current sharply increases at $V= 380~\mathrm{\mu V}$ by approximately 25~nA. This is associated with quasiparticle-quasiparticle tunneling at a voltage bias $V =(\Delta_{1}+\Delta_{2})/e$ corresponding to the sum of the two superconducting gaps. Above this bias, the junction shows ohmic normal-state tunneling. 

In the subgap region below this voltage bias value, there are a number of small subgap tunneling features revealing deviations from BCS tunneling characteristics. At the lowest voltage biases, the measured current slowly decreases. Dissipation from Josephson oscillations in a lossy radio-frequency (RF) environment can give rise to subgap features when the junction sees finite, real impedance at the Josephson frequency. This effect is likely the cause of the anomalous tunneling observed at small biases corresponding to Josephson oscillations from 2 - 20~GHz and could be reduced by measuring a biased SQUID \cite{Stevenson_tunneling}. 

Subgap current steps at individual gap values can also appear in junctions with finite transparency due to multiple Andreev reflections (MAR) \cite{averin-bardas, liao_2019}. MAR theory expands quasiparticle tunneling to junctions with non-zero electron transmission amplitude, $t$. In this case, a quasiparticle  transmitted through the tunnel barrier can ``Andreev reflect'' off the superconducting interface repeatedly between the two superconducting electrodes. In tunneling characteristics, these multiple reflections give rise to steps in the measured current at biases equal to integer fractions of the individual electrode gaps, $\Delta_{1,2}/n$.  The height of these steps is proportional to the junction transparency $D = |t|^2$. In principle, the MAR theory can be used to model the IV data and extract $D$ and $\Delta_{1,2}$ by matching the features in the theory to the measured IV's. Small gap asymmetry is expected due to differences in film thickness \cite{Court_2008, Meservey_1971}, but self-heating, noise, and the nature of some of these novel junctions limit the resolution of this technique. At $V \approx 190\ \mathrm{\mu V}$, the curve in Fig. \ref{fig:IV_comparison} displays an approximate 1 nA step in the current, which we associate with the superconducting gap of the Al electrodes.

Changing the counter-electrode to an Al/Ti bilayer modifies many aspects of the IV characteristics (see the solid curve in Fig. \ref{fig:IV_comparison}). The onset of quasiparticle-quasiparticle tunneling occurs at a reduced voltage bias such that the sum of the superconducting gaps $\Delta_1 + \Delta_2 = 310~\mathrm{\mu eV}$. Because the fabrication of the first layer for all of our devices is identical, the reduction in the sum of the gaps is attributed to reduction of the gap in the top electrode (see Table \ref{tab:table1}), implying that the addition of a thin Ti layer reduces the gap of the structure by nearly $100~\mathrm{\mu eV}$. Steps at $220~\mathrm{\mu eV}$ and $105~\mathrm{\mu eV}$ (see arrows in Fig.~\ref{fig:IV_comparison}) match these proposed gap values. Another striking feature of the IV data for the Al/Ti bilayer junctions is the gradual increase in the current by about 15~nA in the subgap region between $100~\mathrm{\mu eV}$ and $300~\mathrm{\mu eV}$, which is significantly more than the other junctions.

From the MAR theory, the increased current suggests the junction barrier has higher transparency which could lead to greater qubit depolarization rates. Both Al and O can interdiffuse with Ti layers \cite{AlTiO_diffusion}, potentially introducing significant changes to the top electrode or the junction barrier \cite{sapphire_ti}.

Attempting to reduce this subgap current, new devices were fabricated with an additional thin AlO$_y$ disordered layer between the Al and Ti layers. The oxygen doses for this process were less than the dose to form the junction tunnel barrier. For the lowest dose of 0.3~$\mathrm{mBar \cdot min}$, the IV characteristics resemble Al/AlO$\mathrm{_x}$/Al/Ti junctions with no added disorder. However, when the oxygen dose is increased to 0.6~$\mathrm{mBar \cdot min}$, the sum of the gaps and low junction transparency of the initial Al/AlO$\mathrm{_x}$/Al junctions are recovered (see green curve of Fig. \ref{fig:IV_comparison}). 
Further IV characterization of junctions with disorder doses between 0.3 $\mathrm{mBar \cdot min}$ and 0.6 $\mathrm{mBar \cdot min}$ is still required to determine if this bilayer technique can be used to fabricate junctions with both a low transparency and the large desired gap asymmetry ($|\Delta_1 - \Delta_2| > 20~\mu \mathrm{eV}$). Table \ref{tab:table1} shows the extracted top electrode gaps  for all of the measured junctions.

\section{Lifetime Measurements} \label{sec:lifetime}
We fabricated three separate 3D transmons to investigate how these different junctions perform as qubits. The fabricated devices consisted of one Al/AlO$\mathrm{_x}$/Al transmon with no Ti layer, one Al/AlO$\mathrm{_x}$/Al/Ti transmon with a nominally clean interface between the Al and Ti, and one Al/AlO$\mathrm{_x}$/Al/AlO$\mathrm{_y}$/Ti transmon with an oxygen dose of $0.6\ \mathrm{mbar} \cdot \mathrm{min}$ between the Al and Ti depositions. The Al/AlO$\mathrm{_x}$/Al/Ti transmon was fabricated using the Dolan bridge process and had frequency $f_{ge} = 5.1~\mathrm{GHz}$ while the Al/AlO$\mathrm{_x}$/Al and  Al/AlO$\mathrm{_x}$/Al/AlO$\mathrm{_y}$/Ti transmons were fabricated using the Manhattan process and had $f_{ge} = 2.7~\mathrm{GHz}, 4.0~\mathrm{GHz}$ respectively.

\begin{figure}
    \centering
    \includegraphics[width=3.4in]{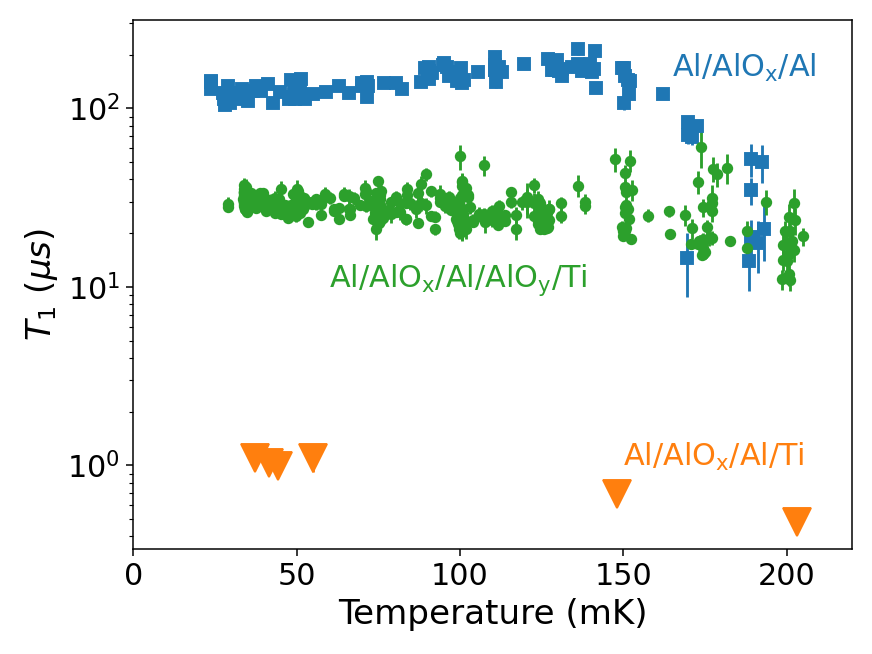}
    \caption{Repeated $T_1$ measurements of three different 3D transmon qubits versus temperature. The squares show data of a standard Al/AlO$_x$/Al qubit. The triangle data had a bilayer of Al/Ti as the counter-electrode. The circle data had an additional thin disordered AlO$_y$ ($0.6~\mathrm{mbar \cdot min}$ O$_2$ dose) in between the Al and Ti counter-electrode. The fluctuations for each qubit are approximately 10\% of the respective average $T_1$ time. 
    }
    \label{fig:T1_vs_dose}
\end{figure}

The standard Al/AlO$\mathrm{_x}$/Al transmon with no Ti performed the best with a mean energy relaxation time $\overline{T}_1 = 134~\mathrm{\mu s}$ at base temperature. When the temperature of the fridge was increased to $T = 140$~mK, the relaxation time increased to $\overline{T}_1 = 210~\mathrm{\mu s}$ before decreasing at higher temperatures (see Fig. \ref{fig:T1_vs_dose}). This temperature dependence of $T_1$ is consistent with previous results \cite{zhang_2020}. 

The Al/AlO$\mathrm{_x}$/Al/Ti transmon with no additional AlO$\mathrm{_y}$ barrier had the worst lifetime with $\overline{T}_1 = 1~\mathrm{\mu s}$. This device's coherence was significantly degraded from the standard Al/AlO$\mathrm{_x}$/Al devices and the DC IV's showed a large amount of subgap current. From these observations, the junction barrier may be compromised by this specific fabrication procedure potentially due to the ability of Al, Ti, and O to diffuse between layers changing material properties of the electrode \cite{AlTiO_diffusion}. To attempt to remove this issue, the next device implemented an AlO$_\mathrm{y}$ barrier in between the Al and Ti bilayer.

The Al/AlO$\mathrm{_x}$/Al/AlO$\mathrm{_y}$/Ti transmon with the additional barrier had a relaxation time with $\overline{T}_1 = 32~\mathrm{\mu s}$. The addition of the disordered layer not only reduced the current in the sub-gap region of the IV's  (Fig. \ref{fig:IV_comparison}), but also increased the $T_{1}$ when compared to the Al/Ti bilayer with no disorder. $T_{1}$ was still degraded compared to the 
Al/AlO$\mathrm{_x}$/Al transmon, but further measurements are necessary to determine device-to-device variability of the disorder bilayered transmons.

Finally, to test the loss of the Ti film, a device with quarter-wave coplanar waveguide resonators made from a 200~nm thick Ti film was fabricated. The low-temperature single photon internal quality factors of these resonators were measured to be $Q_i \simeq 3000$ resulting in an internal photon lifetime of $\kappa^{-1} = 100$~ns, a value even smaller than our transmon qubits.   

Unlike Al, Ti does not form a self-limiting oxide \cite{AlTiO_diffusion, Ti_alloys}. This could result in a greater amount of dielectric loss or surface resistance loss and thus a greater rate of energy decay as observed in the devices with Ti presented here.


\section{Conclusion}
By depositing a thin Ti film on top of Al counter-electrode, the Al is proximitized and reduced the superconducting gap from approximately $\Delta = 200~\mathrm{\mu eV}$ to $100~\mu V$ for the Al/Ti bilayer. This bilayer was used to create low critical current asymmetric superconducting gap Josephson junctions. DC differential conductance measurements at 20~mK were performed to compare these junctions to standard Al/AlO$_\mathrm{x}$/Al junctions as well as devices with additional AlO$_\mathrm{y}$ disorder placed in between the Al and Ti layer. 3D transmon qubits with  added Ti  had reduced energy relaxation times    with the Al/Ti device having the smallest $T_1 = 1.0~\mathrm{\mu s}$. The transmon fabricated with AlO$_\mathrm{y}$ disorder between the Al and Ti showed intermediate lifetimes up to $T_1 = 32~\mu $s. Future work could explore burying the Ti under other metal layers to reduce oxidation, fabricating asymmetric junctions with other superconducting materials, as well as investigating the material composition used for these devices.

\section*{Acknowledgements} We would like to thank Chris Lobb, Alan Kramer, Jon Cripe, Edwin Supple, and Brian Gorman for useful discussions on the fabrication, measurement, and interpretation of results.


\bibliography{references}

\end{document}